\documentstyle[12pt]{article}

\def\be {\begin{equation}}
\def\ee {\end{equation}}
\def\ba{\begin{array}}
\def\ea{\end{array}}

\def\tr{\mbox{Tr}}
\newcommand{\aaa}{A_\alpha}
\newcommand{\ha}{H_\alpha}
\newcommand{\gab}{g_{\alpha\beta}}
\newcommand{\gba}{g_{\beta\alpha}}
\newcommand{\la}{\Lambda_\alpha}

\newcommand{\calh}{{\cal H}}
\newcommand{\ra}{\rho_\alpha}
\newcommand{\ca}{{\cal A}}
\newcommand{\dia}{\mbox{diag} }
\newtheorem{definition}{Definition}
\newtheorem{algorithm}{PDP Algorithm}
\def\lqq{\lq\lq}
\def\rqq{\rq\rq}

\begin{document}
\begin{titlepage}
\today          \hfill
\begin{center}
\hfill BiBoS 683/4/95
\vskip .5in

{\large \bf  Theory of Events}
\footnote{Talk given by the second author at the
Conference on \lqq Nonlinear, Dissipative,
Irreversible Quantum Systems\rqq,
Clausthal 1994.}
\vskip .50in

\vskip .5in
Ph.~Blanchard${}^\flat$ \ and\ A.~Jadczyk${}^\sharp$\footnote{
e-mail: ajad@ift.uni.wroc.pl}

{\em ${}^\flat$ Faculty of Physics and BiBoS,
University of Bielefeld\\
Universit\"atstr. 25,
D-33615 Bielefeld\\
${}^\sharp$ Institute of Theoretical Physics,
University of Wroc{\l}aw\\
Pl. Maxa Borna 9,
PL-50 204 Wroc{\l}aw}
\end{center}

\vskip .5in

\begin{abstract}
We review what we call \lqq event-enhanced formalism\rqq of quantum
theory. In this approach we explicitly assume classical nature of
events.  Given a quantum system, that is coupled
to a classical one by a suitable coupling, classical events are being
triggered. The trigerring process is partly random and partly deterministic.
Within this new approach one can modelize real experimental events,
including pointer readings of measuring devices.
Our theory gives, for the first time, a unique algorithm that can be used
for computer generation of experimental runs with individual quantum objects.
\end{abstract}
\end{titlepage}
\newpage
\section{Introduction}

We will talk about \lqq theory of events\rqq. To be honest we
should allow for the adjective \lqq phenomenological\rqq. We will explain
later our reasons for this restraint. This new theory
enhances and extends the standard quantum formalism.
It provides a solution to the quantum measurement problem.
The usual formalism of quantum theory fails in this respect. Let us look,
for instance, into a recent book on the subject,
\lqq The interpretation of quantum theory\rqq \cite{omnes}. There we can
see both
the difficulties as well as the methods that attempt to overcome them.
We disagree with the optimism shared by many, perhaps by a majority
of quantum physicists. They seem to believe that the
problem is already solved, or almost solved. They use a magic spell,
and at present the magic spell that is supposed to dissolve the
problems is \lqq decoherence\rqq.
It is true that there are new ideas and new results in the decoherence
approach. But these
results did not quite solve the problem. Real--world--events, in particular
pointer readings of measuring apparata, have never be obtained within this
approach. Decoherence does not tell us yet
how to programm a computer to simulate such events.
A physicist, a human being, must intervene to decide what to decohere
and how to
decohere. Which basis is to be distinguished. What must be neglected and what
must not? Which limit to take?
That necessity of a human intervention is not a surprise.
The standard quantum formalism simply has no resources that
can be called for when we wish to {\em derive} the basic postulates about
measurements
and probabilities. These postulates are repeated in all textbooks.
They are never {\sl derived}. The usual probabilistic interpretation of
quantum theory is {\em postulated} from outside. It is not deduced
from within the  formalism. That is rather unsatisfactory.
We want to believe that quantum theory is fundamental, but its
interpretation is so arbitray! Must it be so?

Many physicists would oppose. They disagree with such a criticism.
They see that
quantum theory is good, is excellent, because it gives excellent results.
But there are other voices too. We
like to recall John Bell's opinion on this matter. He has studied the
subject rather deeply.
He emphasized it repeatedly \cite{bell89,bell90}:
our problems with quantum measurements have a source.
The reason is
that the very concept of \lqq measurement\rqq {\em can not even be precisely
defined} within the
standard formalism. That is also our opinion. But not only we share his
criticism. We also
propose a way out that is new.\\
Our solution does not involve hidden variables
(but we like to joke that the standard quantum state vector can be
considered as a hidden variable). Our reasoning goes as follows:\\
First, we point out the
reason {\em why} \lqq measurement\rqq  could not be defined within the
standard approach.
It is true that the standard
formalism of quantum theory has many sophisticated tools:
it has Hilbert spaces, wave
vectors, operators, spectral measures, POV measures. But it has no
place for \lqq events\rqq . What constitutes an event? The only candidate
for an event that we can think of is change of a
quantum state vector. But how do we observe state vectors? We can not
see them directly. We were taught by Bohr and Heisenberg
that any observation will disturb
a quantum state. Well, unless the state is already known to us,
then we can try
to be clever and not to disturb it. But how can we know the state? We need a
theory, that would help us to unswer these questions. We are
proposing such a theory. We extend
the standard formalism. We do it in a minimal way: just enough to
accomodate classical events. We add explicitly
a classical part to the quantum part, and we couple classical to the quantum.
Then we define
\lqq experiments\rqq and \lqq measurements\rqq  within the so extended
formalism.
We can show that the standard postulates concerning measurements -- in fact,
in an enhanced and refined form --
can be derived instead of being postulated.

This \lqq event enhanced
quantum theory\rqq, as we call it,  gives experimental predictions that are
stronger than
those obtained from the standard theory. The new theory gives answers to more
experimental questions than the old one. It provides
algorithms for numerical simulations of experimental time series given by
experiments with single quantum systems. In particular this new theory is
falsifiable. But our programm is not yet complete.
Our theory is based on an explicit
selection of a classical subsystem. How to select what is classical? If we
want to be on a save side as much as possible, or as long as possible, then
we will shift \lqq classical\rqq into the observer's mind. But will we
be save then? For how long? Soon we will need to extend our theory and to
include a theory of mind and a theory
of knowledge. That necessity will face us anyhow, pehaps even soon.
But it is not
clear that the cut must reside that far from
the ordinary physics.
For many practical
applications the measuring  apparatus itself, or its relevant part,
can be considered classical.
We need to derive
such a splitting into classical and quantum from some clear principles.
At present we do not know what these principles are, we can only guess.

At the
present stage placement of the split is indeed phenomenological,
and the coupling is phenomenological too. Both are simple to handle and easy
to describe
in our formalism. But where to put the Heisenberg's cut -- that is arbitrary
to
some extent. Perhaps we need not worry too much? Perhaps relativity
of the split is a new feature that will remain with us. We do not know.
That is why we call our theory \lqq phenomenological\rqq. But we would like
to stress that the standard, orthodox, pure quantum theory is not better in
this respect. In fact, it is much worse. It is not even able to define
what measurement is. It is not even a phenomenological theory. In fact,
strictly speaking, it is not even a theory. It is partly an art, and
that needs an artist. In this case
it needs a physicist with his human experience and with his human intuition.
Suppose we have a problem that needs quantum theory for its solution.
Then our physicist, guided by his intuition,  will replace the
problem at hand by another problem, that can be handled. After that, guided
by his experience, he will compute Green's function or whatsoever to
get formulas out of this other problem. Finally, guided by his previous
experience and by his intuition, he will interpret the formulas that he got,
and he will predict
some numbers for the experiment.\\ That job
can not be left to
a computing machine in an unmanned space--craft. We may feel proud that
we are that necessary, that we can not be replaced by machines.
But would it not be better if we could spare our creativity
for inventing new theories rather than spending it unnecessarily
for application of the old ones?

Our theory is better in this
respect.
Once we have chosen a model -- then reality,
with all its events as they happen in time, can be simulated by a
sufficiently powerful digital computer.

\section{The formalism}

Let us sketch the mathematical framework. To define
events, we introduce a classical system $C$, and possible events will be
identified with changes of a (pure) state of $C$. Let us consider
the simplest
situation corresponding to a finite set of possible events. If
necessary, we can handle infinite dimensional
generalizations of this framework. The space of states of the classical
system, denoted by ${\cal S}_c$, has $m$
states, labelled by $\alpha=1,\ldots,m$. These are the pure states of $C$.
They correspond to possible results of single observations of $C$.\\
Statistical states of $C$ are probability measures on ${\cal S}_c$ -- in
our case just sequences $p_\alpha\geq 0, \sum_\alpha p_\alpha=1$. They
describe ensambles of observations.\\ We
will also need the algebra of (complex) observables of $C$. This will be
the algebra $\ca_c$ of complex functions on ${\cal S}_c$ -- in our case
just sequences $f_\alpha, \alpha =1,\ldots,m$ of complex numbers.\\ It is
convenient to use Hilbert space language even for the description of that
simple classical system. Thus we introduce an $m$-dimensional Hilbert space
$\calh_c$ with a fixed basis, and we realize $\ca_c$ as the algebra of
diagonal matrices $F=\dia(f_1,\ldots,f_m)$.\\ Statistical states of $C$ are
then diagonal density matrices $\dia(p_1,\ldots,p_m)$, and pure states of
$C$ are vectors of the fixed basis of $\calh_c$.\\ Events are ordered
pairs of pure states $\alpha\rightarrow\beta$, $\alpha\neq\beta$. Each
event can thus be represented by an $m\times m$ matrix with $1$ at the
$(\alpha,\beta)$ entry, zero otherwise. There are $m^2-m$ possible events.\\
Statistical states are concerned with ensembles, while pure states and events
concern individual systems.\\ The simplest classical system is a yes--no
counter. It has only two distinct pure states. Its algebra of observables
consists of $2\times 2$ diagonal matrices.

We now come to the quantum system. Here we use the standard description.\\
Let $Q$ be the quantum system whose
bounded observables are from the algebra
$\ca_q$ of bounded operators on a Hilbert space $\calh_q$.
Its pure states are unit vectors in $\calh_q$;
proportional vectors describe the same quantum state. Statistical
states of $Q$ are given by non--negative density matrices ${\hat\rho}$,  with
$\tr ({\hat\rho})=1$. Then pure states can be identified with those density
matrices that are idempotent ${\hat\rho}^2={\hat\rho}$, i.e. with
one--dimensional orthogonal projections.

Let us now consider the total system $T=Q\times C$. Later on we will
define \lqq experiment\rqq as a coupling of $C$ to $Q$. That coupling will
take place within $T$.\\ First, let us consider
statistical description, only after that we shall discuss dynamics and
coupling of the two systems. \\
For the algebra $\ca_t$ of observables of $T$
we take the tensor product of algebras of observables of $Q$ and $C$:
$\ca_t=\ca_q\otimes\ca_c$. It acts on the tensor product
$\calh_q\otimes\calh_c=\oplus_{\alpha=1}^m\calh_\alpha$, where
$\calh_\alpha\approx\calh_q.$ Thus
$\ca_t$ can be thought of as algebra of {\em diagonal} $m\times m$
matrices $A=(a_{\alpha\beta})$, whose entries are quantum operators:
$a_{\alpha\alpha}\in \ca_q$, $a_{\alpha\beta}=0$ for $\alpha\neq\beta$.\\
The classical and quantum algebras are then subalgebras of $\ca_t$;
$\ca_c$ is realized by putting $a_{\alpha\alpha}=f_\alpha I$, while
$\ca_q$ is realized by choosing $a_{\alpha\beta}=a\delta_{\alpha\beta}$.\\
Statistical states of $Q\times C$ are given by $m\times m$ diagonal matrices
$\rho=\dia(\rho_1,\ldots,\rho_m)$ whose entries are positive operators
on $\calh_q$, with the normalization $\tr (\rho)=\sum_\alpha\tr
(\ra)=1$. Tracing over $C$ or $Q$ produces the effective states of $Q$ and
$C$ respectively: ${\hat\rho}=\sum_\alpha \ra$, $p_\alpha=\tr (\ra )$.\\
Duality between observables and states is provided
by the expectation value $<A>_\rho=\sum_\alpha \tr (\aaa\ra)$.

We consider now dynamics. Quantum dynamics, when no information is
transferred from $Q$ to $C$, is described by Hamiltonians $\ha$,
that may depend on the actual state of $C$ (as indicated by the index
$\alpha$). They may also depend explicitly on time.
We will use matrix notation and write $H=\dia(\ha)$.
Now take the classical system. It is discrete here.
Thus it can not have continuous time
dynamics of its own.

Now we come to the crucial point -- our main invention.
A {\em coupling} of $Q$ to $C$ is specified by a matrix
$V=(\gab)$, with $g_{\alpha\alpha}=0$. To transfer  information
from $Q$ to $C$ we need a non--Hamiltonian term which provides
a completely positive (CP) coupling. We propose to consider
couplings for which the evolution
equation for observables and for states is given by the Lindblad  form:
\be
{\dot A}=i[H,A]+{\cal E}\left(V^\star AV\right)-{1\over2}\{\Lambda,A\},
\ee
\be
{\dot \rho}=-i[H,\rho]+{\cal E}(V\rho V^\star)-{1\over2}\{\Lambda,\rho\},
\ee
where
${\cal E}:(A_{\alpha\beta})\mapsto \dia (A_{\alpha\alpha})$ is the
conditional expectation
onto the diagonal subalgebra given by the diagonal projection, and
\be
\Lambda={\cal E}\left(V^\star V\right).
\ee
We can also write it down in a form not involving ${\cal E}$:
\be
{\dot A}=i[H,A]+\sum_{\alpha\neq\beta}V_{[\beta\alpha]}^\star
AV_{[\beta\alpha]}-{1\over2}\{\Lambda,A\},
\ee
with $\Lambda$ given by
\be
\Lambda=\sum_{\alpha\neq\beta}V_{[\beta\alpha]}^\star V_{[\beta\alpha]},
\ee
and where $V_{[\alpha\beta]}$ denotes the matrix that has only one
non--zero entry, namely $\gab$ at the $\alpha$ row and $\beta$ column.
Expanding the matrix form we have:
\be
{\dot A}_\alpha=i[\ha,\aaa]+\sum_\beta \gba^\star
A_\beta \gba - {1\over2}\{\la,\aaa\},\label{eq:lioua}
\ee
\be
{\dot \rho}_\alpha=-i[\ha,\ra]+\sum_\beta \gab
\rho_\beta \gab^\star - {1\over2}\{\la,\ra\},\label{eq:liour}
\ee
where
\be
\la=\sum_\beta \gba^\star \gba.
\ee
Again, the operators $\gab$ can be allowed to depend explicitly
on time.

Following \cite{ja94c} we now define {\em experiment} and
{\em measurement}:

\begin{definition}
An {\bf experiment} is a CP coupling between a quantum and a classical
system. One observes then the classical system and attempts to learn
from it about characteristics of state and of dynamics of the quantum
system.
\end{definition}

\begin{definition}
A {\bf measurement} is an experiment that is used for a particular
purpose: for determining values, or statistical distribution of values,
of given physical quantities.
\end{definition}

The universe that we know, including us, the observers, can be considered
as an \lqq experiment\rqq. That point is discussed in \cite{blaja94pl}.

\section{The algorithm for events}

The definition of experiment above is concerned with the {\em conditions}
that define it. We will now describe the
algorithm that simulates a typical {\em run} of a given experiment.
That algorithm can be uniquely derived from the above formalism. One
then gets the correct statistics by averaging over individual runs.

Let us first make a side but important remark.
In practical situations it is rather easy to decide what constitutes $Q$,
what constitutes $C$ and how to write down the coupling. Then, if
necessary, we enlarge $Q$, and we shift $C$ towards more macroscopic
and/or more classical. The new point of view that we propose
allows us to consider our whole Universe as \lq experiment\rq\ in which
we are witnesses and participants of one particular run. Then one can
ask: {\em what is the true} $C$? We do not know yet. Perhaps it
has something to do with massless particles, with light, with photon
detections. But perhaps we should not postpone any asking questions
that are hard for a physicist: what is Knowledge and what is Mind?

Back to the main subject.
It can be shown that there is a unique
Markov process taking place in pure states of the total system that gives,
after averaging over individual runs, time evolution of statistical states
as described by Eq. (\ref{eq:liour}). That process is piecewise
deterministic -- we call it PDP.
Continuous evolution is interspersed with random jumps.
Here it is:
\begin{algorithm}
Let us assume a fixed, sufficiently small, time step $dt$.
Suppose that at time $t$ the
system is described by a quantum state vector $\psi$ and a classical
state $\alpha$. Compute the scalar product
$\lambda(\psi,\alpha)=<\psi,\la\,\psi>$.
Toss dies and choose a uniform random number $p\in [0,1]$. Jump
if $p<\lambda(\psi,\alpha)dt$. Otherwise not jump.
When jumping, toss dies and change $\alpha\rightarrow
\beta$ with probability $p_{\alpha\rightarrow\beta}=
\Vert\gba\psi\Vert^2/\lambda(\psi,\alpha)$, and change
$\psi\rightarrow\gba\psi/
\Vert\gba\psi\Vert$. If not jumping, change

$$
\psi \rightarrow
{{\exp \{-i\ha dt-{1\over2}\la dt\}\psi}\over{\Vert
\exp \{-i\ha dt-{1\over2}\la dt\}\psi\Vert}} ,\quad
t \rightarrow  t+dt.
$$ Repeat the steps.\footnote{There are
several methods available for
efficient computation of the exponential for $dt$ small enough --
cf. Ref. \cite{raedt87}.}
\end{algorithm}
For derivation and for a proof of uniqueness of the algorithm -- see
\cite{blaja94pl}.
Our algorithm resembles that known in quantum
optics as Wave Function Monte Carlo \cite{carm93,dal92,molmer93,dum92,gard92}.
But there is an important difference: we did not guess our process. We
derived it from
M.H.A. Davis' mathematical theory of PDP processes \cite{davis93}.
We were also able to prove its uniqueness. That
could not be achieved before. In fact, there is no uniqueness without an
explicit introduction of a classical
system. Ten years ago Diosi \cite{dio1} (see also \cite{dio2}) introduced
\lqq orthojump\rqq
 process as a canonical solution to a master equation. His solution although
canonical is not unique -- unless one makes Hilbert spaces corresponding
to different experimental situations orthogonal -- as it is the case with our
$\calh_\alpha$-s.

We have mentioned in the beginning that our theory is falsifiable. Indeed,
the PDP algorithm predicts time series of experimental events. They are
changes of state of $C$. The continuous evolution between these events
is affected be the coupling -- it is {\em non--unitary} and {\em non--linear}.
Its non--linearity depends on the copupling. Several examples have been
already worked out. Some
of them, including a SQUID--tank model, can be found in \cite{blaja93c}.
A cloud chamber model and its relation to GRW spontaneous localization
models \cite{grw86} have been worked out in \cite{ja94b}.

\section*{Acknowledgment(s)}

One of us (A.J.) would like to thank A. von Humboldt Foundation for
the support.

\end{document}